\documentclass[conference]{IEEEtran}
\IEEEoverridecommandlockouts
\usepackage{cite}
\usepackage{comment}
\usepackage{amsmath,amssymb,amsfonts}
\usepackage{algorithmic}
\usepackage{graphicx}
\usepackage{textcomp}
\usepackage{xcolor}
\usepackage{booktabs}
\usepackage{multirow}
\usepackage{pifont}
\usepackage{float}
\def\BibTeX{{\rm B\kern-.05em{\sc i\kern-.025em b}\kern-.08em
    T\kern-.1667em\lower.7ex\hbox{E}\kern-.125emX}}
\begin{document}

\title{\LARGE \bf
Context-Aware Adaptive Shared Control for Magnetically-Driven Bimanual Dexterous Micromanipulation\\
}

\author{\IEEEauthorblockN{Yongchen Wang*, Kangyi Lu*, Lan Wei, Dandan Zhang}
\thanks{*Equal contributions. All authors are with the Department of Bioengineering, Imperial-X, Imperial College London.}
}

\maketitle

\begin{abstract}
Magnetically actuated robots provide a promising untethered platform for navigation in confined environments, enabling biological studies and targeted micro-delivery. However, dexterous manipulation in complex structures remains challenging. While single-arm magnetic actuation suffices for simple transport, steering through tortuous or bifurcating channels demands coordinated control of multiple magnetic sources to generate the torques required for precise rotation and directional guidance. Bimanual teleoperation enables such dexterous steering but imposes high cognitive demands, as operators must handle the nonlinear dynamics of magnetic actuation while coordinating two robotic manipulators. 
To address these limitations, we propose Bi-CAST, a context-aware adaptive shared control framework for bimanual magnetic micromanipulation. A multimodal network fuses spatio-temporal visual features, spatial risk metrics, and historical states to continuously adjust the control authority of each manipulator in real time. In parallel, a bidirectional haptic interface integrates force-based intent recognition with risk-aware guidance, enabling force feedback to provide a continuous channel for dynamic human-machine authority negotiation.
We validate the framework through user studies with eight participants performing three navigation tasks of increasing complexity in a vascular phantom. Compared with fixed authority and discrete switching baselines, Bi-CAST achieves up to 76.6\% reduction in collisions, 25.9\% improvement in trajectory smoothness, and 44.4\% lower NASA-TLX workload, while delivering the fastest task completion times.

\end{abstract}

\section{INTRODUCTION}



Magnetic micromanipulation has emerged as a promising paradigm for minimally invasive biomedical applications, with the long-term goal of enabling untethered robotic agents to navigate complex vascular networks for tasks such as targeted drug delivery and intravascular intervention~\cite{landers2025clinically, lin2024magnetic, zhang2026situ}. This approach is attractive due to its wireless control, compatibility with miniaturized devices, and non-contact force generation through biological tissue. However, reliable navigation through confined and branching vascular structures remains challenging, as it demands high precision and continuous corrective control under nonlinear magnetic dynamics and constrained visual feedback from microscopic imaging.

\begin{figure}[t]
	\centering
	\includegraphics[width = 1\hsize]{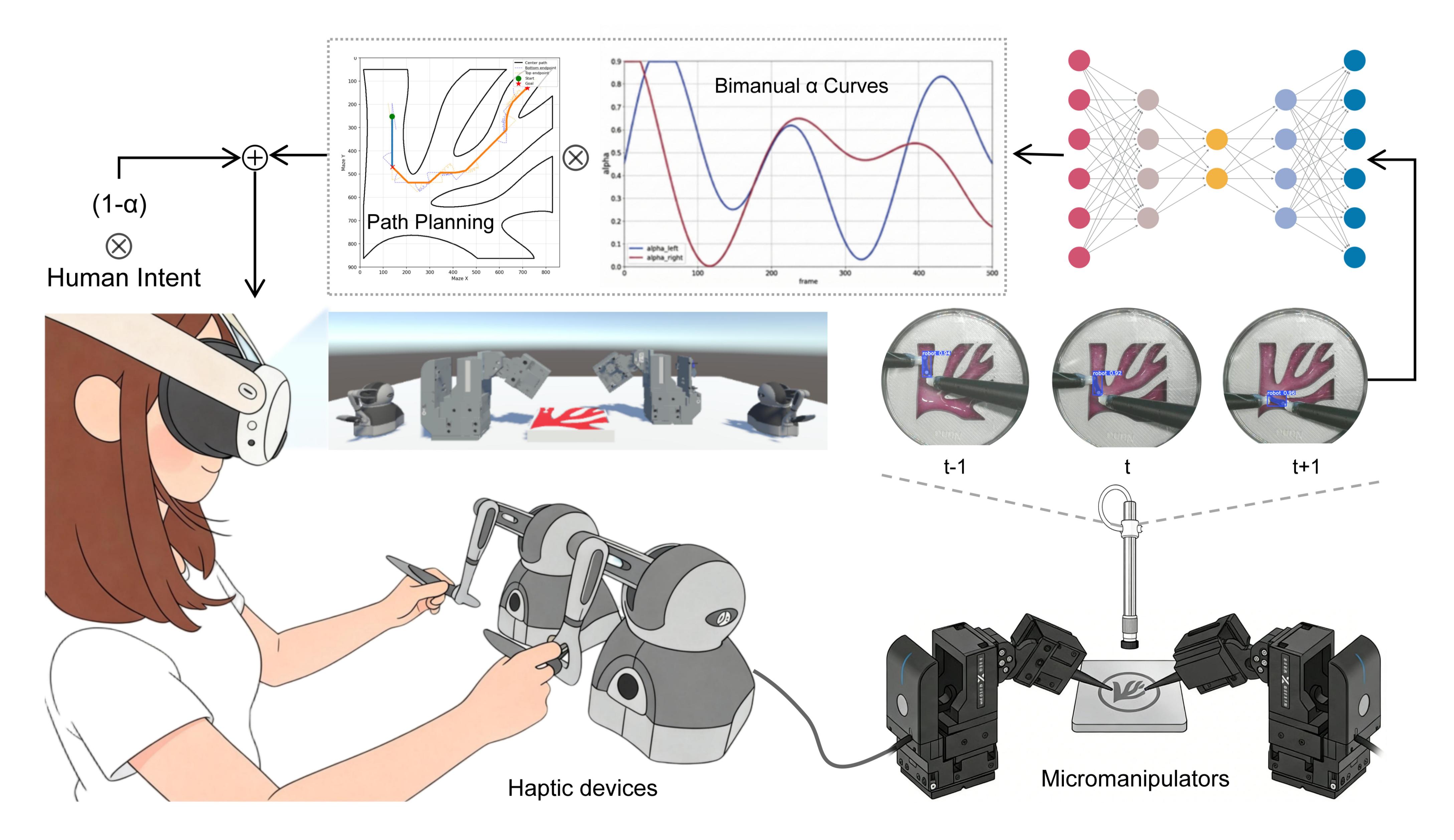}
    \vspace{-0.8cm}
	\caption{Overview of the bimanual shared micromanipulation framework, integrating human intent signals and autonomous path planning via a dynamic blending factor $\alpha$ predicted by a spatio-temporal network based on sequential video frames, enabling continuous and independent authority allocation. }
	\label{fig:system overview}
    \vspace{-14pt}
\end{figure}

Existing control strategies for magnetic miniature robot navigation generally fall into two categories: direct teleoperation and autonomous control. Teleoperation enables expert operators to exploit human intuition and decision-making to guide the robot through complex environments \cite{zhang2022teleoperation}, but it requires sustained attention and precise motor coordination. These challenges are further exacerbated by limited spatial awareness, making prolonged teleoperation cognitively demanding and prone to errors \cite{duygu2025real}. Fully autonomous approaches aim to reduce operator burden through automatic planning and execution, yet often struggle to adapt to unexpected environmental variations and lack the transparency required in safety-critical medical contexts \cite{liu2024autonomous}.


Shared control provides a promising compromise between these paradigms by combining human decision-making with robotic assistance \cite{payne2020shared}. In shared-control systems, authority over the robot's motion is dynamically distributed between the human operator and an autonomous controller \cite{li2019shared}. This collaboration allows the operator to retain high-level decision-making while the robot assists with low-level motion stabilization, improving trajectory precision, reducing cognitive load, and enhancing safety by preventing potentially harmful actions while preserving operator oversight. Haptic feedback can further enhance this collaboration by conveying environmental interactions through tactile cues, improving situational awareness beyond what vision alone can provide \cite{lin2023tims,riaziat2024investigating}. However, existing haptic-enhanced shared-control approaches for magnetic miniature robots predominantly rely on discrete, pre-defined blending strategies or binary mode switching, resulting in abrupt control transitions and motion discontinuities without continuous adaptation to operational context~\cite{ferro2024experimental}.

Beyond discontinuous blending, existing frameworks often lack real-time spatial awareness, treating human-machine interaction as unidirectional command execution rather than continuous intent negotiation \cite{chen2020supervised}, limiting their ability to provide smooth, conflict-free assistance during complex manipulation tasks. Moreover, most prior approaches have focused on single-arm magnetic actuation, which offers limited steering authority in tortuous or bifurcating channels. In such geometries, the trailing end of the robot responds passively to the applied field, leading to lateral deviation and wall collision risks. Bimanual magnetic actuation addresses this by enabling two manipulators to independently control the fields at the robot's head and tail, allowing coordinated rotation and more precise steering through complex branches. However, this dual-arm configuration substantially increases control dimensionality, placing greater cognitive demands on the operator and making precise manual teleoperation considerably more challenging.

To address these limitations, we propose Bi-CAST, a context-aware adaptive shared-control framework for bimanual magnetic micromanipulation. Unlike existing approaches that rely on discrete mode switching, the proposed framework models authority negotiation as a continuous process, enabling smooth transitions between human and autonomous control \cite{zhang2022human}. To estimate authority weights in real time, we develop a multimodal network that integrates spatio-temporal visual features, bilateral force-sensitive resistor (FSR) signals for user intent detection, and spatial safety metrics describing the surrounding environment. These inputs are fused using a Transformer encoder, followed by two independent linear heads that predict the authority weight $\alpha$ for each manipulator. To ensure stable interaction dynamics, a multi-step chunk prediction mechanism forecasts authority weights over several future time steps and applies a smoothness constraint to discourage abrupt transitions. In addition, a bidirectional haptic interface integrates force-based intent recognition with risk-aware guidance, allowing the operator to continuously negotiate control authority with the learned policy while receiving tactile feedback about environmental interactions. The main contributions of this work are summarized as follows:
\begin{itemize}
\item 
We propose a multimodal authority negotiation framework that moves beyond discrete mode switching with continuous adaptive blending, enabling smooth and stable shared control.

\item 
We introduce an adaptive shared-control framework that independently scales the control authority of two manipulators, enabling coordinated steering of a millirobot through branching vascular geometries.

\item 
We develop a haptic interaction framework that integrates operator intent recognition with risk-aware guidance, establishing force feedback as a bidirectional communication channel for dynamic authority negotiation.
\end{itemize}

\section{RELATED WORK}

\subsection{Control Strategies for Magnetic Millirobot Navigation}

Teleoperation leverages expert intuition for magnetic miniature robot control. Liu et al. proposed a computer-aided system mapping user inputs to robot actions \cite{liu2024computer}, and Riaziat et al. demonstrated that haptic feedback improves task performance by over 40\% under degraded visual conditions \cite{riaziat2024investigating}. However, simultaneously coordinating two robotic arms significantly increases the operator's mental workload and degrades control precision, particularly in tasks requiring asymmetric or tightly coupled bimanual motions \cite{zhang2025scheduling}.

To reduce operator burden, autonomous navigation methods have been actively explored, including robust 3D path following under fluid disturbances~\cite{qi2024robust}, navigation in uncertain dynamic environments~\cite{tian2025automatic}, and reinforcement learning-based 3D positional control~\cite{abbasi2024autonomous}. Despite these advances, fully autonomous systems remain limited in their ability to generalize across diverse anatomical geometries and cannot incorporate real-time operator judgment when unexpected situations arise \cite{liu2024autonomous}. This tension motivates shared control strategies that bridge operator intuition and robotic precision.


\subsection{Shared Control and Authority Allocation}

In the context of micromanipulation, shared control addresses the limitations of pure teleoperation by distributing task execution between a human operator and an autonomous agent, where a blending ratio $\alpha$ determines the level of automation (Table \ref{tab:related_work}). Recent work has explored haptic-assisted teleoperation for magnetic microrobots. Ferro et al. demonstrated 65\% gains in positioning accuracy with haptic target guidance~\cite{ferro2024experimental}, and Raphalen et al. combined Control Barrier Functions with haptic cues for safe microrobot pair navigation~\cite{raphalen2025haptic}. However, these approaches rely on predefined discrete authority modes. Similarly, Mao et al. proposed a DRL-based binary switching framework that reduces task time but introduces abrupt transitions \cite{mao2025deep}. To mitigate such discontinuities, context-aware allocation methods have been explored. Li et al. employed bidirectional trust metrics with receding horizon optimization \cite{li2024reconciling}, and Zhong et al. used obstacle-proximity-based sigmoid functions \cite{zhong2026adaptive}.

However, these methods rely on single-metric heuristics and fixed analytical rules, without accounting for the increased dimensionality of bimanual actuation. Our framework addresses this gap by fusing multimodal context, including spatio-temporal visual features, occlusion uncertainty, and haptic intent, to enable continuous adaptive authority allocation for bimanual magnetic micromanipulation.


\begin{table}[t]
\centering
\footnotesize
\caption{Comparison of shared control  in micromanipulation.}
\vspace{-8pt}
\label{tab:related_work}
\renewcommand{\arraystretch}{0.5}
\setlength{\tabcolsep}{4pt}
\begin{tabular}{l c c c c}
\toprule
\textbf{Work} & \textbf{Bimanual} & \textbf{Shared Ctrl} & \textbf{Ctx-Aware} & \textbf{Adaptive} \\
\midrule
Raphalen et al.~\cite{raphalen2025haptic}   & \ding{55} & \ding{51} & $\triangle$ & \ding{55} \\
Ferro et al.~\cite{ferro2024experimental}   & \ding{55} & \ding{51} & $\triangle$ & \ding{55} \\
Mao \& Zhang~\cite{mao2025deep}             & \ding{55} & \ding{51} & $\triangle$ & $\triangle$ \\
Li et al.~\cite{li2024reconciling}          & \ding{55} & \ding{51} & \ding{51} & \ding{51} \\
Zhong et al.~\cite{zhong2026adaptive}       & \ding{55} & \ding{51} & \ding{51} & \ding{51} \\
\midrule
\textbf{Bi-CAST (Ours)}                    & \ding{51} & \ding{51} & \ding{51} & \ding{51} \\
\bottomrule
\end{tabular}
\vspace{4pt}
\par\footnotesize\textbf{Note:} \ding{51}~Yes, \enspace \ding{55}~No, \enspace $\triangle$~Partial.
\vspace{-12pt}
\end{table}

 \section{Methodology}
 \subsection{System Overview}




The proposed Bi-CAST framework continuously and independently allocates control authority to each manipulator in real time. To eliminate the motion discontinuities of traditional mode-switching, a Transformer-based multimodal network fuses spatio-temporal visual features, spatial safety metrics, and bilateral operator intent (captured via FSR sensors) to regress independent authority parameters $\alpha$ for each arm. These parameters continuously modulate the mapping between operator input and manipulator motion, seamlessly adapting human control authority to the operational context.

As shown in Fig. \ref{fig:system overview}, the operator controls the two micromanipulators via haptic devices, with each end-effector displacement mapped to the corresponding manipulator’s translational motion. A top-down microscope captures live images of the millirobot, which are tracked and streamed to a Unity-based digital twin, providing immersive visualization that enhances situational awareness and manipulation precision. Operator intent can be expressed through applied force on the haptic
devices, which the network interprets to adjust the balance between human and autonomous guidance.


\begin{figure}[t]
	\centering
	\includegraphics[width = 1\hsize]{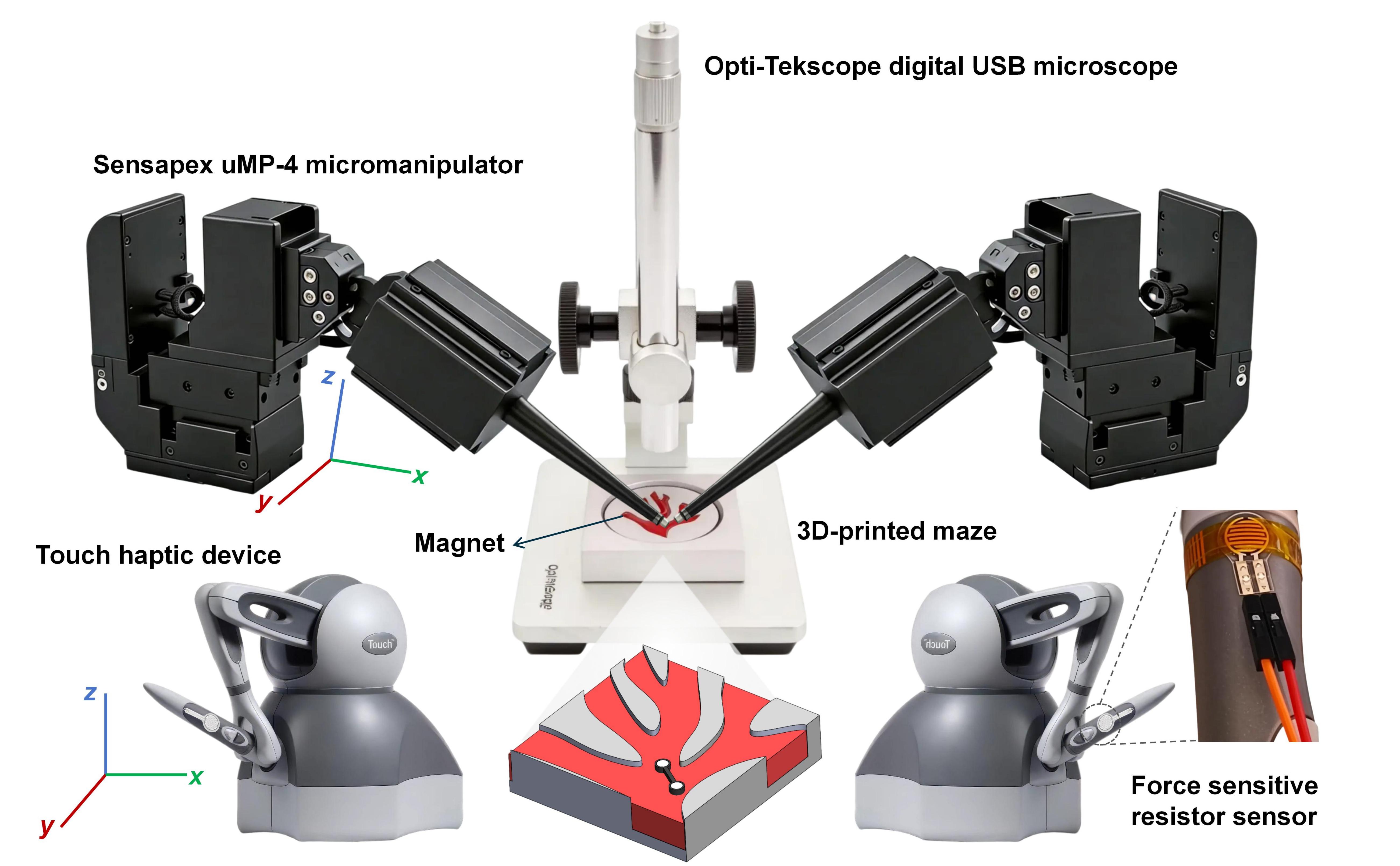}
	\caption{Overview of the bimanual micromanipulation platform. Two micromanipulators are arranged in an opposing configuration, each tipped with a magnet for millirobot actuation. The operator controls the system through two  haptic devices equipped with FSR sensors for user intent capture. A digital microscope provides real-time imaging.}
	\label{fig:hardware}
    \vspace{-14pt}
\end{figure}

\subsection{Hardware System}
Each manipulator provides a 20 mm positioning range along the X, Y, and Z axes, as well as linear motion along the shaft (D axis) via a closed-loop piezoelectric motor mechanism. Custom end-effector mounts, fabricated using FDM 3D printing, rigidly hold an axially magnetized permanent magnet ($\varnothing$1 mm × 2 mm) at each tip. The opposing arrangement of the magnets generates a spatially controlled magnetic field, enabling precise actuation of the millimeter-scale robot.

The millirobot was fabricated using an SLA 3D-printed mold, with a pair of NdFeB magnets ($\varnothing$1 mm × 1 mm) embedded at both ends in an anti-parallel axial magnetization configuration, resulting in a total body length of 4 mm. Navigation experiments were conducted within a vascular phantom featuring a tree-like branching topology, where a main trunk bifurcates into multiple subsidiary channels of reduced cross-section, replicating the hierarchical architecture of physiological vasculature.



Operator hand motions are captured by the Geomagic Touch devices and forwarded to the manipulator controller, which sends position commands to the uMp-4 units via the Sensapex SDK. To quantify operator intent, two FSR sensors are mounted on each Geomagic Touch to measure grip force, providing a continuous signal of user intent. When the millirobot approaches vessel boundaries, repulsive force feedback is rendered on the devices to alert the operator. Simultaneously, the Opti-Tekscope digital USB microscope (1600 × 1200 px, 30 Hz) provides a continuous 2D image stream for online millirobot localization, supplying input to both the visual tracker and the authority negotiation network.



\subsection{Bidirectional Haptic Interaction}
\textbf{\textit{1) Haptic Feedback Rendering}}: Haptic devices are widely employed in teleoperation to render environmental interaction forces to the operator, enhancing situational awareness and reducing reliance on visual feedback alone \cite{patel2022haptic}. Two complementary haptic feedback channels are rendered on each Geomagic Touch throughout the navigation task. First, a repulsive force is generated when the millirobot approaches vessel boundaries:
\begin{equation}
F_{rep} = \begin{cases} k_{rep}\left(\frac{1}{d} - \frac{1}{d_0}\right)\frac{1}{d^2}\hat{n}, & d \leq d_0 \\ 0, & d > d_0 \end{cases}
\end{equation}
where $d_0$ is the influence threshold, $k_{rep}$ is the repulsive gain, and $\hat{n}$ is the unit vector pointing away from the nearest wall. Second, an attractive guidance force steers the operator toward the reference path:
\begin{equation}
F_{guide} = k_{guide} \cdot \mathbf{e}_{path},
\end{equation}
where ${k}_{guide}$ is the guidance stiffness gain and ${e}_{path}$ is the deviation vector from the millirobot's current position to the nearest point on the reference path. The two channels are combined as:
\begin{equation}
F_{haptic} = F_{rep} + (1 - \alpha) \cdot F_{guide},
\end{equation}
where the guidance force fades naturally as $\alpha$
increases, ensuring haptic guidance diminishes as the autonomous system takes greater authority.

\textbf{\textit{2) Operator Intent Recognition via FSR}}: Operator intent is captured through two FSR sensors mounted on each Geomagic Touch, whose analog voltage signals are read via an Arduino microcontroller. Sensors were calibrated by applying a series of known weights to establish the empirical mapping between ADC readings and actual force values.

To account for inter-user variability in grip force, a personalized calibration is performed before each session. The operator is first asked to hold the device with a natural resting grip, recording the baseline force $F_{baseline}$, and then with a deliberate override grip, recording $F_{override}$. The normalized intent signal is then defined as:
\begin{equation}
I(t) = \frac{F_{FSR}(t) - F_{baseline}}{F_{override} - F_{baseline}} \in [0, 1].
\end{equation}
To suppress transient noise, a sliding window average with ${w = 3}$ is applied:\begin{equation}
\bar{I}(t) = \frac{1}{w}\sum_{k=0}^{w-1} I(t-k).
\end{equation}

\begin{figure*}[t]
    \centering
    \includegraphics[width=0.9\linewidth]{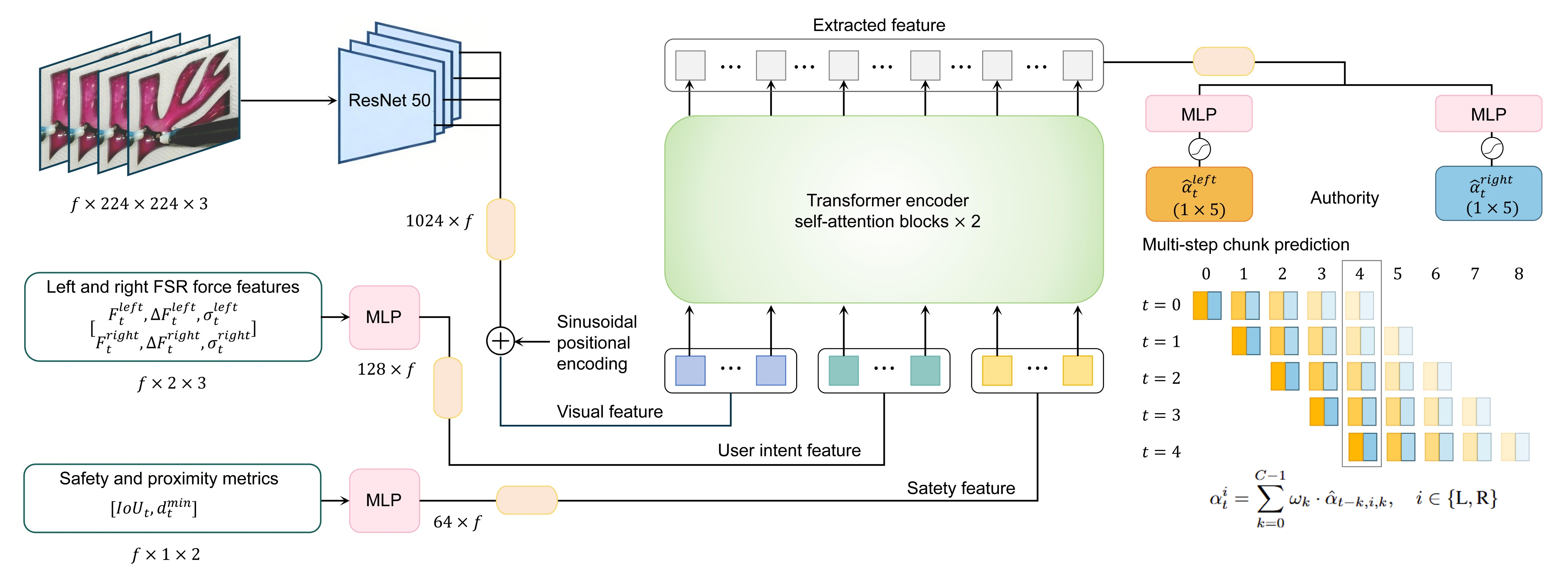}
    \vspace{-10pt}
    \caption{Architecture of the multimodal authority prediction network. Three input streams (RGB frames, bilateral FSR force features, and safety metrics) are encoded and fused through a Transformer encoder with eight self-attention blocks. The network outputs bilateral authority chunks of five future steps, which are causally aggregated via exponential temporal weighting: $\alpha_t^i = \sum_{k=0}^{C-1} \omega_k \hat{\alpha}_{t-k,i,k}$.}
    \label{fig:structure}
    \vspace{-0.6cm}
\end{figure*}

\subsection{Digital Twin-Enhanced Immersive Teleoperation}

To enable intuitive real-time monitoring of the micromanipulation task, an immersive digital twin framework is constructed within Unity to mirror the physical workspace. The framework operates along two concurrent paths: a forward path that transmits control commands from the haptic devices to the micromanipulators, and a feedback path that relays visual and positional information back to the operator. To handle occlusion caused by the manipulator tips during operation, a registration-then-tracking strategy is employed. The millirobot is first registered in the initial frame by matching its physical geometry, and its position is subsequently maintained through a visual tracker across frames, ensuring robust localization even under partial occlusion. The tracked position is continuously streamed to Unity via TCP to update the virtual scene, which is then rendered on a Meta Quest 3 head-mounted display. Through the headset, the operator can observe the vascular phantom and the robot position, ensuring full scene awareness even under physical occlusion.

\subsection{Context-Aware Authority Negotiation Algorithm}

Our bimanual shared control framework blends operator input with autonomous guidance for each arm. The command issued to each manipulator is formulated as:
\begin{equation}
\mathbf{u}_i(t) = \alpha_i(t) \cdot \mathbf{u}_{robot,i}(t) + (1 - \alpha_i(t)) \cdot \mathbf{u}_{human,i}(t)
\end{equation}
where $i \in \{L, R\}$, $\mathbf{u}_{human,i}(t)$ denotes the operator's hand motion for arm $i$ captured by the corresponding Geomagic Touch, $\mathbf{u}_{robot,i}(t)$ denotes the autonomous navigation command for arm $i$, and $\alpha_i(t)$ is the estimated authority parameter for each arm. When $\alpha_i = 0$, the operator retains full control of arm $i$; as $\alpha_i$ increases, greater authority is delegated to the autonomous system. To preserve a minimum level of human supervisory control in this safety-critical context, $\alpha_i$ is bounded within $[0, 0.9]$.

The autonomous navigation command $\mathbf{u}_{robot,i}(t)$ is generated by a path planner that operates on a safety costmap constructed from the vascular phantom's free-space mask. A distance transform is applied to the traversable region to compute per-cell traversal costs, penalizing proximity to vessel walls. An A* search is then performed on this costmap to generate a centerline path from the current robot position to the designated target \cite{mansfield2020topological}. Modelling the robot as a fixed-length rigid rod, independent waypoint sequences for both manipulator tips are derived from the centerline path via rigid-body kinematics, with dual-arm collision detection applied to ensure that arms do not overlap throughout the trajectory.

Rather than computing $\alpha$ from a single heuristic signal, we train a multimodal network to regress it from operational context. As illustrated in Fig.~\ref{fig:structure}, the network receives three input streams over a temporal window of $f$ frames. The visual stream processes $f$ consecutive RGB frames ($224 \times 224 \times 3$) through a ResNet-50 backbone, producing a $2048 \times f$ feature sequence that is linearly projected to $1024 \times f$ and augmented with positional embeddings. The user intent stream encodes bilateral FSR force features $[F_t, \Delta F_t, \sigma_t]$ (raw force, derivative, and variance) for both hands, projected by an MLP into a $128 \times f$ representation. The safety stream encodes $[\text{IoU}_t, d_t^{\text{min}}]$ (IoU and minimal wall distances) through a separate MLP, yielding a $64 \times f$ feature sequence. The visual and intent features are concatenated and fed together with the safety features into an eight-layer Transformer encoder.

To overcome the motion discontinuities of prior discrete switching approaches, a multi-step chunk prediction mechanism is adopted~\cite{zhao2023learning}. Rather than regressing $\alpha$ at a single time step, the MLP head simultaneously forecasts bilateral authority chunks $\hat{\alpha}_{t,i,k}$ for $k \in \{0,\dots,C-1\}$ and $i \in \{\text{L},\text{R}\}$, where $C$ denotes the chunk size, with each prediction bounded by a sigmoid activation:
\begin{equation}
    \hat{\alpha}_{t,i,k} = 0.9 \cdot \sigma(\text{logit}_{t,i,k}), \quad i \in \{\text{L}, \text{R}\}
\end{equation}
to ensure the operator retains at least 10\% control authority in this safety-critical context. The final authority at time $t$ is obtained by causally aggregating overlapping chunk predictions that target the same step:
\begin{equation}
    \alpha_t^i = \sum_{k=0}^{C-1} \omega_k \cdot \hat{\alpha}_{t-k,i,k}, \quad i \in \{\text{L}, \text{R}\}
\end{equation}
The training objective is defined as:
\begin{equation}
    \mathcal{L} = \sum_{i \in \{L,R\}} w_i \left[ \mathcal{L}_{\text{Huber}}^{i} + \lambda_s \sum_{k=0}^{C-2}(\hat{\alpha}_{k+1}^{i} - \hat{\alpha}_k^{i})^2 \right] + \lambda_1 \sum |\theta|,
\end{equation}
where $w_i$ denotes the arm-specific loss weight with $w_L = 0.6$ and $w_R = 0.4$, determined via Bayesian hyperparameter search using the Optuna framework~\cite{akiba2019optuna}. $\mathcal{L}_{\text{Huber}}^{i}$ is the Huber loss between predicted and ground-truth $\alpha$ values for each arm $i \in \{L, R\}$, the smoothness term with $\lambda_s = 0.1$ penalizes abrupt changes between consecutive predicted authority values within each chunk, and the last term is L1 regularization over model parameters $\theta$.

\section{EXPERIMENTS AND RESULTS}

\subsection{Experimental Setup}
\begin{figure}[t]
    \centering
    \includegraphics[width=\linewidth]{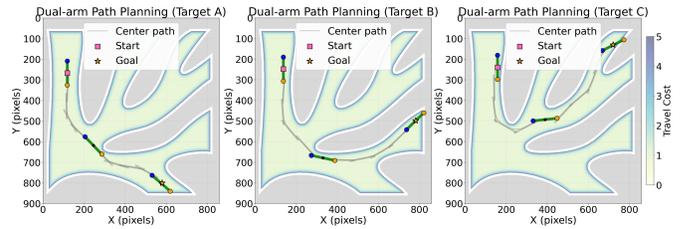}
    \vspace{-0.8cm}
    \caption{Dual-arm path planning across three tasks of increasing complexity: Target A (one bifurcation), Target B (additional bifurcations), and Target C (multiple bifurcations with narrowing channels). The heatmap indicates distance-transform-based travel cost, and overlaid poses show the millirobot's orientation changes along each path.}
    \label{fig:task_def_path_planning}
    \vspace{-0.3cm}
\end{figure}

\textbf{\textit{1) Task Design}}: Shared-control navigation tasks were conducted in a 3D-printed vascular phantom with a tree-like branching topology, in which a main trunk bifurcates into multiple subsidiary channels to replicate the hierarchical structure of vasculature. The phantom was filled with silicone oil dyed red (density 0.98 g/cm$^3$, viscosity 1000 cSt) to simulate the viscous fluidic environment of vessels. Three navigation scenarios of increasing complexity were designed to evaluate performance across varying levels of task difficulty (Fig.~\ref{fig:task_def_path_planning}). Task 1 (Easy) requires navigation to Target A through a short path with one bifurcation and moderate turning angles. Task 2 (Medium) extends to Target B along a longer path with additional bifurcation points and larger turning angles. Task 3 (Hard) navigates to Target C through the most complex route, involving multiple bifurcations, sharp turns, and progressively narrowing channels. A trial is considered successful if the millirobot reaches the designated target with five or fewer wall contacts.

\textit{\textbf{2) Participants and Study Protocol}}: A within-subject user study was conducted with eight participants (6 males, 2 females, aged 22–27). All participants were unfamiliar with the specific experimental tasks, although two had prior experience with teleoperation and mixed-reality devices. Before the formal experiments, each participant completed a standardized training session to familiarize themselves with the system. Four control conditions were evaluated:\\
\indent\textbf{Manual:} Pure human bimanual teleoperation without autonomous assistance.\\
\indent\textbf{Fixed authority:} A constant authority level ($\alpha = 0.5$), serving as the shared-control baseline.\\
\indent\textbf{Discrete switching:} A discrete authority strategy that maintains $\alpha = 0.7$ during normal navigation and performs a hard switch to full manual control ($\alpha = 0$) when the robot-to-goal distance falls below 5mm.\\
\indent\textbf{Proposed method:} The context-aware adaptive shared-control framework proposed in this work.

Each participant completed one trial per condition for each of the three tasks, resulting in a total of 96 trials (8 participants $\times$ 4 conditions $\times$ 3 tasks). Condition order was counterbalanced across participants to minimise learning effects.

\begin{figure}[t]
    \centering
    \includegraphics[width=\linewidth]{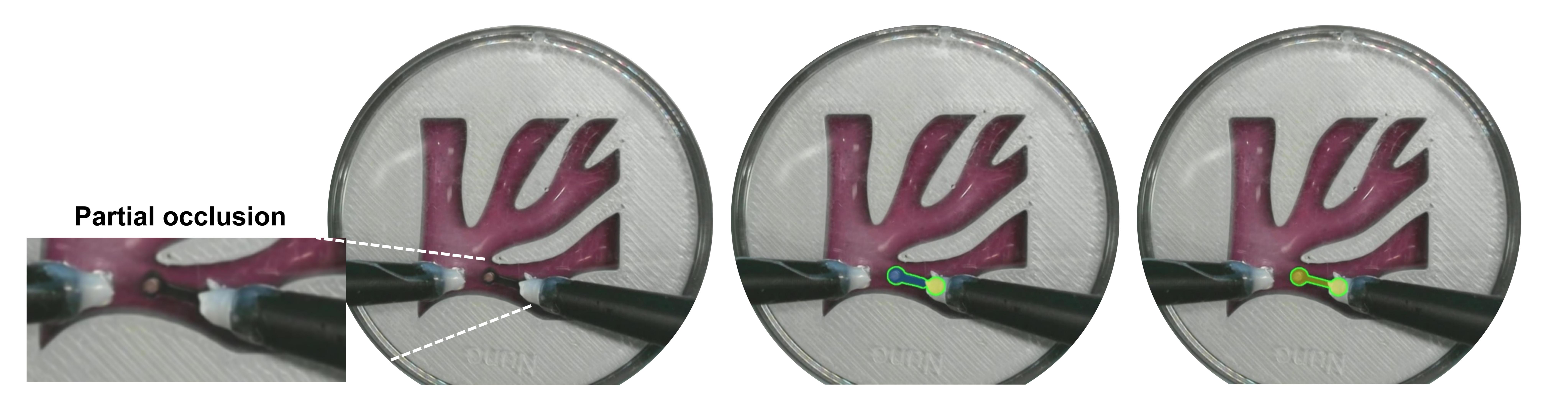}
    \vspace{-0.8cm}
    \caption{Millirobot localization under partial occlusion. (a) Raw microscope image with the manipulator tip occluding the millirobot. (b) YOLO segmentation mask overlaid with tracking result, visualizing IoU. (c) Robust localization maintained via frame-to-frame registration-based tracking.}
    \label{fig:partial_occlusion}
    \vspace{-16pt}
\end{figure}

\begin{table*}[t]
\centering
\caption{Comparison of different network architectures for bilateral authority prediction.}
\label{tab:network_comparison}
\renewcommand{\arraystretch}{0.85}
\vspace{-0.2cm}
\setlength{\tabcolsep}{3pt}
\begin{tabular*}{\textwidth}
{@{\extracolsep{\fill}} l cc cc cc cc}
\toprule
\multicolumn{1}{c}{\multirow{2}{*}{\textbf{Method}}} & \multicolumn{2}{c}{\textbf{MAE $\downarrow$}} & \multicolumn{2}{c}{\textbf{RMSE $\downarrow$}} & \multicolumn{2}{c}{\textbf{R$^2$ $\uparrow$}} & \multicolumn{2}{c}{\textbf{Smoothness $\downarrow$}} \\
\cmidrule(lr){2-3} \cmidrule(lr){4-5} \cmidrule(lr){6-7} \cmidrule(lr){8-9}
& \textbf{L} & \textbf{R} & \textbf{L} & \textbf{R} & \textbf{L} & \textbf{R} & \textbf{L} & \textbf{R} \\
\midrule
3-layer CNN              & 0.2467 & 0.2366 & 0.2826 & 0.2664 & $-$0.051 & $-$0.0031 & 0.4934 & 0.4671 \\
ResNet-18                & 0.1307 & 0.1033 & 0.1813 & 0.1400 & 0.5675 & 0.7232 & 0.3195 & 0.3173 \\
ResNet-50                & 0.1498 & 0.1077 & 0.1982 & 0.1370 & 0.4833 & 0.7347 & 0.4079 & 0.2861 \\
TSM ResNet-18            & 0.1099 & 0.1199 & 0.1363 & 0.1370 & 0.5833 & 0.6347 & 0.3079 & 0.3861 \\
R2Plus1D                 & 0.1608 & 0.1036 & 0.2107 & 0.1244 & 0.4157 & 0.7813 & 0.4984 & 0.3889 \\
MViT                     & 0.1437 &  0.1004 & 0.2002 & 0.1324 & 0.4743 & 0.7526 & 0.4978 & 0.2890 \\
Vision Transformer       & 0.1327 & 0.1014 & 0.1941 & 0.1376 & 0.5042 & 0.7324 & 0.3910 & 0.3276 \\
\midrule
\textbf{Bi-CAST}         & \textbf{0.0318} & \textbf{0.0304} & \textbf{0.0383} & \textbf{0.0356} & \textbf{0.9619} & \textbf{0.9726} & \textbf{0.2593} & \textbf{0.2034} \\
\bottomrule
\end{tabular*}
\vspace{1pt}
\par\footnotesize\textbf{Note:} Best results in \textbf{bold}. $\downarrow$: lower is better; $\uparrow$: higher is better. L/R denote left/right arm.
\vspace{-12pt}
\end{table*}

\subsection{Datasets}
\textbf{\textit{1) Tracking Dataset}}: An instance segmentation dataset of 1,500 images was constructed from video recordings captured across three navigation tasks. Raw video sequences were first subsampled at every other frame to reduce redundancy. SAM 2.1 Hiera Tiny was then employed to propagate instance segmentation masks throughout each sequence by annotating the target robot in the first frame, eliminating the need for frame-by-frame manual annotation \cite{ravi2024sam}. The resulting masks were exported in YOLO format and split into training, validation, and test sets at a ratio of 8:1:1, which were used to fine-tune a YOLOv8s-seg model \cite{sapkota2025yolo26}.


\textbf{\textit{2) Authority Dataset}}:  To train the authority negotiation network, a dataset of 5,000 annotated frames was constructed from video recordings of the three navigation tasks. Frame-wise authority scores were annotated by three independent researchers with prior experience in robotic teleoperation and shared-control systems. The annotators assigned scores by jointly considering several contextual factors: (i) the millirobot's position along the path, (ii) its distance to the nearest bifurcation, (iii) the curvature radius of the current trajectory, and (iv) the degree of visual occlusion caused by the manipulator tips. Annotations were performed independently to minimise individual bias, and the final authority label for each frame was obtained by averaging the scores across the three annotators.
To preserve temporal continuity and prevent data leakage between splits, recordings from Task~1 and Task~3 were used for training, while recordings from Task~2 were divided into validation and test sets, yielding an overall split ratio of 0.7:0.15:0.15.

\subsection{Implementation Details}
The YOLOv8s-seg model, pre-trained on COCO, was fine-tuned for millirobot instance segmentation. Training was conducted for 30 epochs with a batch size of 64 and an input resolution of $640 \times 640$, with standard data augmentation including HSV jittering, random scaling, and mosaic augmentation. Training used SGD with momentum 0.937, an initial learning rate of 0.01 decayed to 0.001 via cosine annealing.

The authority prediction network was trained for 60 epochs using AdamW with a learning rate of $1\times10^{-4}$ and a batch size of 32. The input window spans $f=4$ consecutive frames, each resized to $224 \times 224$ and encoded by a ResNet-50 backbone pretrained on ImageNet into 2048-dimensional features, followed by a linear projection to 1024 dimensions. Bilateral FSR force features and safety metrics are projected to 128 and 64 dimensions through separate MLPs, respectively. The visual features are augmented with sinusoidal positional embeddings and concatenated with the user intent and safety features as input to an eight-layer Transformer encoder. Two independent MLP heads decode the extracted representation into left and right authority chunks of $C=5$ future steps, each bounded by a scaled sigmoid to guarantee at least 10\% operator authority.
The training loss combines Huber loss, L1 regularization ($\lambda_1 = 1\times10^{-3}$), and a smoothness penalty ($\lambda_s = 0.1$) on consecutive predictions. The final authority is obtained via exponential temporal weighting over the chunk. All models were trained and evaluated on a workstation equipped with two NVIDIA A100 GPUs (80 GB each).

\subsection{Results}
\textbf{\textit{1) Millirobot Tracking Performance}}: The fine-tuned model achieved a box detection precision of 97.03\%, mAP@50 of 99.38\%, and mAP@50-95 of 84.70\%, alongside mask segmentation mAP@50 of 99.38\% and mAP@50-95 of 81.24\%, demonstrating reliable millirobot localization under partial occlusion (Fig.~\ref{fig:partial_occlusion}).

\begin{figure}
    \centering
    \includegraphics[width=0.85\linewidth]{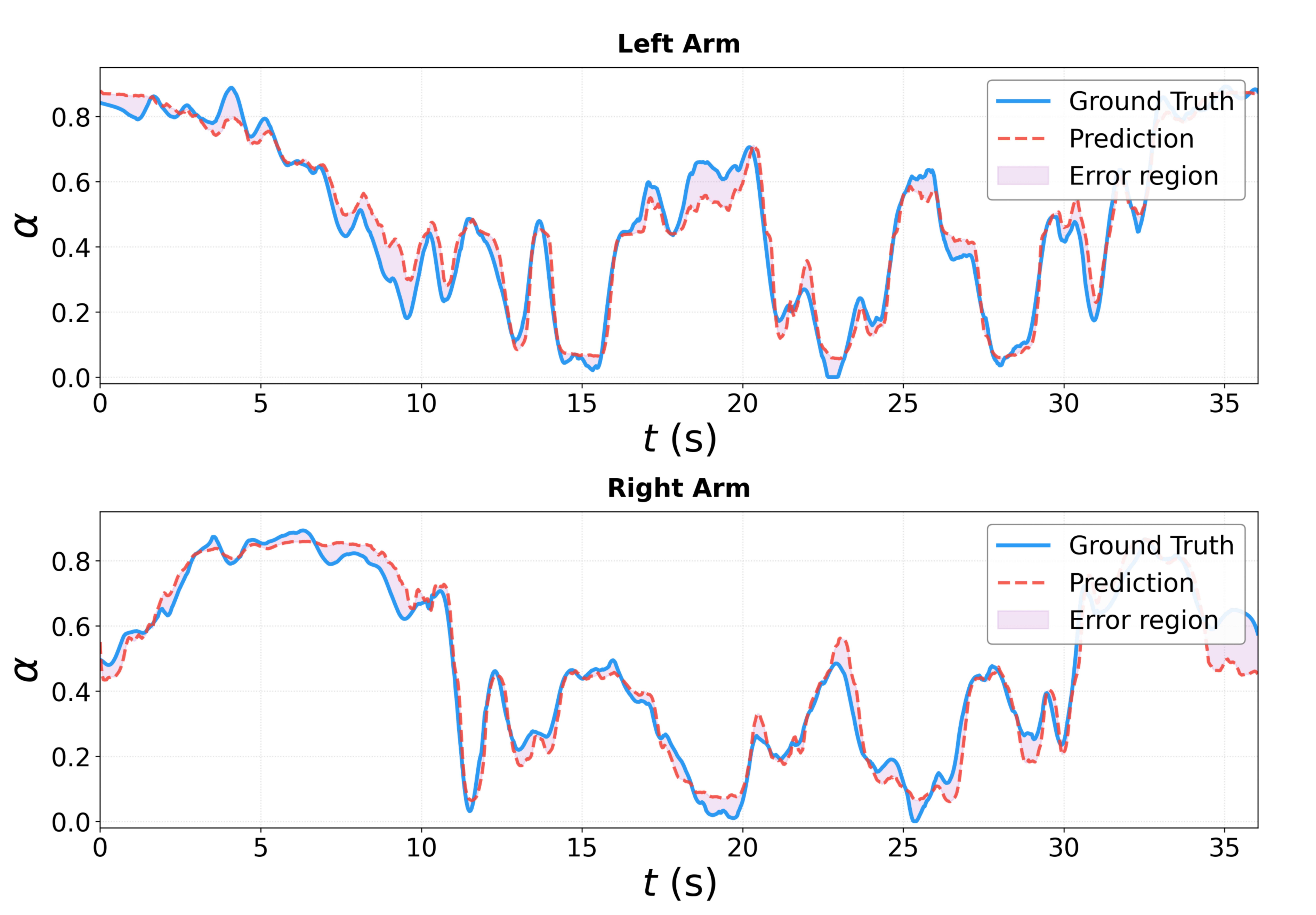}
    \vspace{-8pt}
    \caption{Predicted (red) and ground-truth (blue) authority values $\alpha$ for the left arm (top) and right arm (bottom) over a representative training sequence.}
    \label{fig:alpha_pred}
    \vspace{-16pt}
\end{figure}

\textbf{\textit{2) Authority Allocation Network Comparison}}: We evaluated different network architectures for bilateral authority prediction (Table~\ref{tab:network_comparison}). Image-based backbones (3-layer CNN, ResNet-18, ResNet-50), temporal models (TSM ResNet-18, R2Plus1D), and video Transformers (MViT) mostly yielded R$^2$ values below 0.78, with most architectures showing substantial bilateral imbalance. The Vision Transformer achieved more balanced performance across arms through global self-attention, motivating our Transformer-based design. We selected ResNet-50 as the visual backbone, as its higher-dimensional feature space better complements the Transformer encoder for capturing fine-grained variations. Building on these findings, Bi-CAST employs an eight-layer Transformer encoder to jointly model temporal dependencies across visual, FSR-based user intent signals, and safety modalities. As shown in Fig.~\ref{fig:alpha_pred}, the resulting model closely tracks the ground-truth authority for both arms, capturing bilateral temporal dynamics effectively.

\begin{table}[t]
\centering
\caption{Ablation study on key design components.}
\vspace{-8pt}
\label{tab:ablation}
\renewcommand{\arraystretch}{0.85}
\setlength{\tabcolsep}{3pt}
\begin{tabular*}{\columnwidth}{@{\extracolsep{\fill}} p{1.2cm} p{2.8cm} c c c}
\toprule
& \textbf{Configuration} & \textbf{MAE $\downarrow$} & \textbf{RMSE $\downarrow$} & \textbf{R$^2$ $\uparrow$} \\
\midrule
\multirow{3}{*}{\textit{\shortstack[l]{Visual\\Backbone}}}
& ResNet-18                & 0.0646 & 0.0740 & 0.8929 \\
& ResNet-34                & 0.0364 & 0.0450 & 0.9125 \\
& \textbf{ResNet-50}        & \textbf{0.0311} & \textbf{0.0370} & \textbf{0.9673} \\
\midrule
\multirow{3}{*}{\textit{\shortstack[l]{Input\\Modality}}}
& Vision only              & 0.1196 & 0.1364 & 0.6541 \\
& Vision + FSR             & 0.0609 & 0.0985 & 0.7926 \\
& \textbf{Vision + FSR + Safety} & \textbf{0.0311} & \textbf{0.0370} & \textbf{0.9673} \\
\midrule
\textit{Smooth.} & $\lambda_s = 0$          & 0.0541 & 0.0639 & 0.9358 \\
\midrule
\multirow{3}{*}{\textit{\shortstack[l]{Chunk\\Size}}}
& $C = 1$                  & 0.0530 & 0.1339 & 0.9417 \\
& {$\mathbf{C = 5}$}          & \textbf{0.0311} & \textbf{0.0370} & \textbf{0.9673} \\
& $C = 10$                  & 0.0393 & 0.1465 & 0.8673 \\
\bottomrule
\end{tabular*}
\vspace{1pt}
\par\footnotesize\textbf{Note:} Best in \textbf{bold}. Each group varies one factor with others fixed.
\vspace{-12pt}
\end{table}

\begin{table*}[t]
\centering
\caption{Evaluation metrics: definition, formula, unit, and desired direction.}
\vspace{-8pt}
\label{tab:metrics}
\renewcommand{\arraystretch}{0.8}
\begin{tabular*}{\textwidth}{@{\extracolsep{\fill}} l p{8.5cm} l l l}
\toprule
\textbf{Metric} & \textbf{Definition} & \textbf{Formula} & \textbf{Unit} & \textbf{Dir} \\
\midrule
CT & Total time taken to complete the navigation task from start to target
   & $CT = t_{\text{end}} - t_{\text{start}}$
   & s
   & $\downarrow$ \\
PL & Cumulative distance traveled by the millirobot throughout the trial
   & $PL = \textstyle\sum_{i=1}^{N-1} \|\mathbf{p}_{i+1} - \mathbf{p}_i\|$
   & cm
   & $\downarrow$ \\
Va & Total path length divided by task completion time
   & $Va = PL / CT$
   & cm/s
   & context$^\dagger$ \\
CC & Number of contacts between the robot and vessel walls during a trial
   & --
   & --
   & $\downarrow$ \\
S  & Median of the log-transformed normalized jerk magnitude
   & $S = \mathrm{Median}\!\left(\log_{10}(\phi(T_i))\right)$
   & --
   & $\uparrow$ \\
\bottomrule
\end{tabular*}

\par\footnotesize\textbf{Note:} $\downarrow$ indicates lower is better. $^\dagger$ Higher speed is preferable only when safety metrics remain unchanged.
\vspace{-16pt}
\end{table*}

\textbf{\textit{3) Ablation Study}}: Table~\ref{tab:ablation} presents ablation results on four design components. For the visual backbone, performance scaled with depth, with ResNet-50 achieving R$^2$ of 0.9673 versus 0.8929 for ResNet-18. Each input modality proved essential, as progressively adding FSR-based user intent and safety metrics to the vision stream improved R$^2$ from 0.6541 to 0.9673, confirming the complementary value of multimodal fusion. Disabling the smoothness loss ($\lambda_s = 0$) increased MAE from 0.0311 to 0.0541, confirming that this regularization improves both stability and accuracy. For chunk size, $C = 5$ achieved the best balance, as $C = 1$ suffered from high RMSE due to a lack of temporal context, while $C = 10$ introduced accumulation error over the longer prediction horizon.


\subsection{User Study} 

Upon completing all trials, participants filled out the NASA Task Load Index (NASA-TLX) questionnaire to assess subjective workload across six dimensions: mental demand, physical demand, temporal demand, performance, effort, and frustration \cite{hart2006nasa}. Five objective metrics were also recorded: Completion Time (CT), Path Length (PL), Average Speed (Va), Collision Count (CC), and Smoothness (S), as summarized in Table~\ref{tab:metrics}. Trajectory smoothness was computed from the log-transformed normalized jerk magnitude and min--max normalized to $[0,1]$, where higher values indicate smoother motion. Statistical significance was determined by Friedman tests with Wilcoxon signed-rank post-hoc comparisons and Bonferroni correction.

\begin{table}[t]
\centering
\caption{Comparison of objective performance metrics.}
\label{tab:objective}
\setlength{\tabcolsep}{4pt}
\renewcommand{\arraystretch}{0.85}
\vspace{-0.2cm}
\resizebox{\columnwidth}{!}{%
\begin{tabular}{lcccc}
\toprule
\multicolumn{1}{c}{Metric} & Manual & Fixed authority & Discrete switching & \textbf{Bi-CAST} \\
\midrule
CT $\downarrow$           & $22.530 \pm 7.790$  & $21.580 \pm 8.990$  & $19.350 \pm 4.800$  & $\mathbf{13.680 \pm 4.430}$ \\
PL $\downarrow$           & $1.850 \pm 0.230$   & $1.940 \pm 0.510$   & $1.750 \pm 0.230$   & $\mathbf{1.550 \pm 0.250}$ \\
Va $\uparrow^\dagger$     & $0.088 \pm 0.028$   & $0.100 \pm 0.039$   & $0.093 \pm 0.019$   & $\mathbf{0.118 \pm 0.040}$ \\
S $\uparrow$              & $0.857 \pm 0.166$   & $0.826 \pm 0.179$   & $0.777 \pm 0.225$   & $\mathbf{0.978 \pm 0.027}$ \\
CC $\downarrow$           & $2.240 \pm 1.350$   & $1.920 \pm 1.770$   & $1.650 \pm 1.140$   & $\mathbf{0.450 \pm 0.510}$ \\
\bottomrule
\end{tabular}%
}
\vspace{1pt}
\par\footnotesize\textbf{Note:} All values are reported as mean $\pm$ SD. $\downarrow$: lower is better; $\uparrow$: higher is better. $^\dagger$ Preferable only when safety metrics are unchanged.
\vspace{-16pt}
\end{table}

\subsubsection{\textbf{Quantitative Navigation Performance}}

Table~\ref{tab:objective} compares four control strategies in terms of efficiency (CT, PL, $V_a$), safety (CC), and motion quality (S), aggregated across all tasks. 
Overall, introducing shared control consistently reduced completion time relative to manual, with performance improving from fixed authority (CT = 21.58 s) to discrete switching (CT = 19.35 s) and further to Bi-CAST (CT = 13.68 s). 
However, the two baselines exhibit clear trade-offs. 
Fixed authority slightly increased speed but led to a longer mean path and reduced smoothness compared with manual, indicating that constant blending may bias the motion without improving trajectory quality. 
Discrete switching improved path efficiency and safety compared with fixed authority, but produced the lowest smoothness, consistent with trajectory jitter induced by hard authority transitions. 
In contrast, Bi-CAST achieved balanced improvement across all metrics, delivering the shortest path, the highest speed, and the lowest collision count while also increasing smoothness.
Compared with fixed authority and discrete switching, Bi-CAST reduced CT by 36.6\%/29.3\%, reduced PL by 20.1\%/11.4\%, and reduced CC by 76.6\%/72.7\%, while improving smoothness by 18.4\%/25.9\% (CT/PL/$V_a$/CC: $p<0.05$; S: $p<0.01$), suggesting that continuously varying, context-aware authority negotiation can avoid discontinuities and preserve motion continuity without sacrificing efficiency or safety.

\begin{table}[t]
\centering
\caption{NASA-TLX subjective workload ratings.}
\label{tab:tlx}
\setlength{\tabcolsep}{4pt}
\renewcommand{\arraystretch}{0.85}
\vspace{-0.2cm}
\resizebox{\columnwidth}{!}{%
\begin{tabular}{lcccc}
\toprule
\multicolumn{1}{c}{\textbf{Metric}} & Manual & Fixed authority & Discrete switching & \textbf{Bi-CAST} \\
\midrule
Mental       & $7.04 \pm 2.79$ & $6.21 \pm 2.30$ & $5.17 \pm 1.99$ & $\mathbf{3.58 \pm 1.93}$ \\
Physical     & $7.21 \pm 2.52$ & $6.04 \pm 2.18$ & $5.38 \pm 2.14$ & $\mathbf{3.65 \pm 1.82}$ \\
Temporal     & $6.88 \pm 2.88$ & $5.71 \pm 2.37$ & $5.06 \pm 2.33$ & $\mathbf{3.33 \pm 1.74}$ \\
Performance  & $5.62 \pm 3.32$ & $5.21 \pm 2.67$ & $4.77 \pm 2.48$ & $\mathbf{2.46 \pm 2.00}$ \\
Effort       & $7.04 \pm 2.60$ & $6.00 \pm 2.19$ & $4.92 \pm 1.93$ & $\mathbf{3.33 \pm 1.90}$ \\
Frustration  & $6.58 \pm 3.09$ & $5.71 \pm 2.37$ & $4.92 \pm 1.89$ & $\mathbf{3.04 \pm 1.71}$ \\
\midrule
Overall      & $6.73$          & $5.81$          & $5.04$          & $\mathbf{3.23}$ \\
\bottomrule
\end{tabular}%
}
\vspace{1pt}
\par\footnotesize\textbf{Note:} Mean $\pm$ SD, 0-10 scale (lower is better). Best in \textbf{bold}. \\Overall: unweighted mean of six dimensions. 
\vspace{-12pt}
\end{table}

\begin{figure}[t]
    \centering
    \includegraphics[width=\hsize]{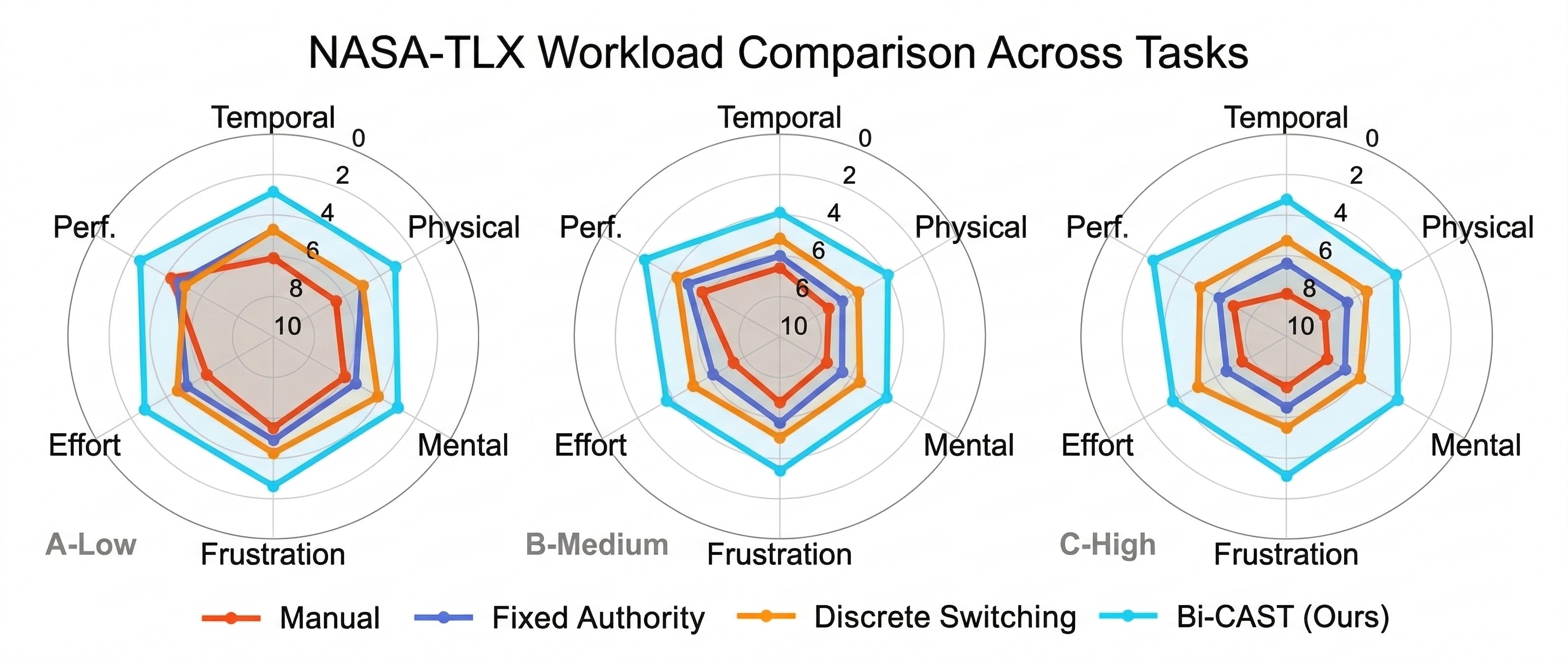}
    \vspace{-0.6cm}
    \caption{NASA-TLX workload comparison across tasks of increasing complexity. Each radar chart shows the six workload dimensions for manual (red), fixed authority (blue), discrete switching (orange), and Bi-CAST (teal).}
    \label{fig:tlx_comparison}
    \vspace{-12pt}
\end{figure}

\subsubsection{\textbf{Subjective Workload}} 

Table~\ref{tab:tlx} presents the NASA-TLX scores for each dimension. All three shared control methods reduced workload compared to manual control, confirming the general benefit of autonomous assistance. Among the shared control strategies, Bi-CAST achieved the lowest workload across all six dimensions. Compared to discrete switching, Bi-CAST further reduced mental demand by 31\% (3.58 vs 5.17), physical demand by 32\% (3.65 vs 5.38), and effort by 32\% (3.33 vs 4.92).
Compared with fixed authority and discrete switching, Bi-CAST reduced mental demand
by 42.4\%/30.8\%, physical demand by 39.6\%/32.2\%, and effort by 44.5\%/32.3\%
($p < 0.05$), while the overall raw TLX score decreased by 44.4\%/35.9\%
($p < 0.01$), suggesting that continuous, context-aware authority adjustment
eliminates the cognitive overhead of abrupt transitions, allowing operators to focus
on task-level decision-making rather than low-level control corrections.



These trends held across all three tasks. Although workload rose with task complexity for all methods, Bi-CAST consistently maintained the lowest scores (Fig.~\ref{fig:tlx_comparison}), demonstrating a clear advantage over all baselines in both objective performance and subjective experience.





\section{CONCLUSIONS}

This paper presented \textbf{Bi-CAST}, a context-aware adaptive shared-control framework for bimanual magnetic micromanipulation that addresses limitations of existing approaches, including discrete mode switching, single-arm actuation, and unidirectional human--machine interaction. A multimodal Transformer fuses spatio-temporal visual features, bilateral FSR-based user intent signals, and spatial safety metrics to predict continuous authority parameters, while a bidirectional haptic interface enables force-based authority negotiation and risk-aware guidance. Experiments on a 3D-printed vascular phantom across three navigation tasks demonstrated strong performance: Bi-CAST achieved bilateral authority prediction with MAE $<0.032$ and $R^2>0.96$. Compared with fixed-authority and discrete-switching baselines, Bi-CAST reduced collisions by up to $76.6\%$, improved trajectory smoothness by $25.9\%$, and lowered NASA-TLX workload by $44.4\%$, demonstrating improvements in both task performance and operator experience.

Future work will extend the current 2D visual tracking and bounding-box perception to 3D state estimation using stereo or multi-view imaging, enabling depth-aware authority allocation based on 3D pose tracking. We also plan to incorporate fluidic disturbance estimation from tracking residuals, allowing the shared control policy to adapt authority weights in response to flow-induced drift and environmental uncertainty.

\bibliographystyle{IEEEtran}
\bibliography{references}

@inproceedings{lin2023tims,
  title={Tims: A tactile internet-based micromanipulation system with haptic guidance for surgical training},
  author={Lin, Jialin and Guo, Xiaoqing and Fan, Wen and Li, Wei and Wang, Yuanyi and Liang, Jiaming and Liu, Jindong and Liu, Weiru and Wei, Lei and Zhang, Dandan},
  booktitle={2023 IEEE/RSJ International Conference on Intelligent Robots and Systems (IROS)},
  pages={5114--5121},
  year={2023},
  organization={IEEE}
}

@article{li2019shared,
  title={Shared control with a novel dynamic authority allocation strategy based on game theory and driving safety field},
  author={Li, Mingjun and Song, Xiaolin and Cao, Haotian and Wang, Jianqiang and Huang, Yanjun and Hu, Chuan and Wang, Hong},
  journal={Mechanical Systems and Signal Processing},
  volume={124},
  pages={199--216},
  year={2019},
  publisher={Elsevier}
}

@article{payne2020shared,
  title={Shared-control robots},
  author={Payne, Christopher J and Vyas, Khushi and Bautista-Salinas, Daniel and Zhang, Dandan and Marcus, Hani J and Yang, Guang-Zhong},
  journal={Neurosurgical Robotics},
  pages={63--79},
  year={2020},
  publisher={Springer}
}

@article{zhang2022teleoperation,
  title={From teleoperation to autonomous robot-assisted microsurgery: A survey},
  author={Zhang, Dandan and Si, Weiyong and Fan, Wen and Guan, Yuan and Yang, Chenguang},
  journal={Machine Intelligence Research},
  volume={19},
  number={4},
  pages={288--306},
  year={2022},
  publisher={Springer}
}

@article{lin2024magnetic,
  title={Magnetic microrobots for in vivo cargo delivery: A review},
  author={Lin, Jialin and Cong, Qingzheng and Zhang, Dandan},
  journal={Micromachines},
  volume={15},
  number={5},
  pages={664},
  year={2024},
  publisher={MDPI}
}

@article{ferro2024experimental,
  title={Experimental evaluation of haptic shared control for multiple electromagnetic untethered microrobots},
  author={Ferro, Marco and Basualdo, Franco N Pinan and Giordano, Paolo Robuffo and Misra, Sarthak and Pacchierotti, Claudio},
  journal={IEEE Transactions on Automation Science and Engineering},
  volume={22},
  pages={8069--8080},
  year={2024},
  publisher={IEEE}
}

@article{zhang2026situ,
  title={In situ mechanostimulation of biohybrid millirobots for enhanced cell functionality and delivery},
  author={Zhang, Jianhua and Bao, Xianqiang and Zhu, Zhou and Zhang, Rongjing and Wang, Chunxiang and Li, Mingtong and Xu, Kaichen and He, Yong and Hutmacher, Dietmar W and Ren, Ziyu and others},
  journal={Science Advances},
  volume={12},
  number={1},
  pages={eadx9616},
  year={2026},
  publisher={American Association for the Advancement of Science}
}

@article{landers2025clinically,
  title={Clinically ready magnetic microrobots for targeted therapies},
  author={Landers, Fabian C and Hertle, Lukas and Pustovalov, Vitaly and Sivakumaran, Derick and Oral, Cagatay M and Brinkmann, Oliver and Meiners, Kirstin and Theiler, Pascal and Gantenbein, Valentin and Veciana, Andrea and others},
  journal={Science},
  volume={390},
  number={6774},
  pages={710--715},
  year={2025},
  publisher={American Association for the Advancement of Science}
}

@inproceedings{mao2025deep,
  title={Deep Reinforcement Learning-Based Semi-Autonomous Control for Magnetic Micro-robot Navigation with Immersive Manipulation},
  author={Mao, Yudong and Zhang, Dandan},
  booktitle={2025 IEEE International Conference on Robotics and Automation (ICRA)},
  pages={9088--9094},
  year={2025},
  organization={IEEE}
}

@article{duygu2025real,
  title={Real-time teleoperation of magnetic force-driven microrobots with a motion model and stable haptic force feedback for micromanipulation},
  author={Duygu, Yasin Cagatay and Xie, Baijun and Zhang, Xiao and Kim, Min Jun and Park, Chung Hyuk},
  journal={Nanotechnology and Precision Engineering},
  volume={8},
  number={2},
  year={2025},
  publisher={AIP Publishing}
}

@inproceedings{hart2006nasa,
  title={NASA-task load index (NASA-TLX); 20 years later},
  author={Hart, Sandra G},
  booktitle={Proceedings of the human factors and ergonomics society annual meeting},
  volume={50},
  number={9},
  pages={904--908},
  year={2006},
  organization={Sage publications Sage CA: Los Angeles, CA}
}

@inproceedings{zhang2022human,
  title={Human-robot shared control for surgical robot based on context-aware sim-to-real adaptation},
  author={Zhang, Dandan and Wu, Zicong and Chen, Junhong and Zhu, Ruiqi and Munawar, Adnan and Xiao, Bo and Guan, Yuan and Su, Hang and Hong, Wuzhou and Guo, Yao and others},
  booktitle={2022 International conference on robotics and automation (ICRA)},
  pages={7694--7700},
  year={2022},
  organization={IEEE}
}

@inproceedings{chen2020supervised,
  title={Supervised semi-autonomous control for surgical robot based on Bayesian optimization},
  author={Chen, Junhong and Zhang, Dandan and Munawar, Adnan and Zhu, Ruiqi and Lo, Benny and Fischer, Gregory S and Yang, Guang-Zhong},
  booktitle={2020 IEEE/RSJ International Conference on Intelligent Robots and Systems (IROS)},
  pages={2943--2949},
  year={2020},
  organization={IEEE}
}

@article{liu2024autonomous,
  title={Autonomous navigation of magnetic microrobots with improved planning and control in complex environments},
  author={Liu, Yueyue and Wang, Haoyu and Wu, Xiaoyu and Qu, Juntian and Liu, Xinyu and Fan, Qigao},
  journal={IEEE Transactions on Automation Science and Engineering},
  volume={22},
  pages={2421--2432},
  year={2024},
  publisher={IEEE}
}

@article{abbasi2024autonomous,
  title={Autonomous 3D positional control of a magnetic microrobot using reinforcement learning},
  author={Abbasi, Sarmad Ahmad and Ahmed, Awais and Noh, Seungmin and Gharamaleki, Nader Latifi and Kim, Seonhyoung and Chowdhury, AM Masum Bulbul and Kim, Jin-young and Pan{\'e}, Salvador and Nelson, Bradley J and Choi, Hongsoo},
  journal={Nature Machine Intelligence},
  volume={6},
  number={1},
  pages={92--105},
  year={2024},
  publisher={Nature Publishing Group UK London}
}

@article{riaziat2024investigating,
  title={Investigating haptic feedback in vision-deficient millirobot telemanipulation},
  author={Riaziat, Naveed D and Erin, Onder and Krieger, Axel and Brown, Jeremy D},
  journal={IEEE robotics and automation letters},
  volume={9},
  number={7},
  pages={6178--6185},
  year={2024},
  publisher={IEEE}
}

@article{li2024reconciling,
  title={Reconciling conflicting intents: Bidirectional trust-based variable autonomy for mobile robots},
  author={Li, Yinglin and Cui, Rongxin and Yan, Weisheng and Zhang, Shi and Yang, Chenguang},
  journal={IEEE Robotics and Automation Letters},
  volume={9},
  number={6},
  pages={5615--5622},
  year={2024},
  publisher={IEEE}
}

@article{zhong2026adaptive,
  title={Adaptive Shared Cascade Navigation Control of Magnetic Microrobots in Unstructured Dynamic Environments},
  author={Zhong, Shihao and Hou, Yaozhen and Zheng, Zhiqiang and Huang, Hen-Wei and Shi, Qing and Huang, Qiang and Fukuda, Toshio and Wang, Huaping},
  journal={IEEE Transactions on Cybernetics},
  year={2026},
  publisher={IEEE}
}

@article{qi2024robust,
  title={Robust 3-D path following control framework for magnetic helical millirobots subject to fluid flow and input saturation},
  author={Qi, Zhaoyang and Cai, Mingxue and Hao, Bo and Cao, Yanfei and Su, Lin and Liu, Xurui and Chan, Kai Fung and Yang, Chenguang and Zhang, Li},
  journal={IEEE transactions on cybernetics},
  volume={54},
  number={12},
  pages={7629--7641},
  year={2024},
  publisher={IEEE}
}

@article{tian2025automatic,
  title={An automatic navigation framework for magnetic fish-like millirobot in uncertain dynamic environments},
  author={Tian, Chengyao and Fan, Xinjian and Jia, Jingzhi and Yang, Zhan and Xie, Hui},
  journal={IEEE Robotics and Automation Letters},
  volume={10},
  number={3},
  pages={2422--2429},
  year={2025},
  publisher={IEEE}
}

@article{liu2024computer,
  title={A Computer-Aided Teleoperation System for Intuitively Controlling the Behavior of a Magnetic Millirobot within a Stomach Phantom},
  author={Liu, Ruomao and Xiang, Yuxuan and Wei, Zihan and Zhang, Jiachen},
  journal={Advanced Intelligent Systems},
  volume={6},
  number={2},
  pages={2300325},
  year={2024},
  publisher={Wiley Online Library}
}

@article{patel2022haptic,
  title={Haptic feedback and force-based teleoperation in surgical robotics},
  author={Patel, Rajni V and Atashzar, S Farokh and Tavakoli, Mahdi},
  journal={Proceedings of the IEEE},
  volume={110},
  number={7},
  pages={1012--1027},
  year={2022},
  publisher={IEEE}
}

@article{zhao2023learning,
  title={Learning fine-grained bimanual manipulation with low-cost hardware},
  author={Zhao, Tony Z and Kumar, Vikash and Levine, Sergey and Finn, Chelsea},
  journal={arXiv preprint arXiv:2304.13705},
  year={2023}
}

@inproceedings{mansfield2020topological,
  title={A Topological Approach to Path Planning for a Magnetic Millirobot},
  author={Mansfield, Ariella and Kularatne, Dhanushka and Steager, Edward and Hsieh, M Ani},
  booktitle={2020 IEEE/RSJ International Conference on Intelligent Robots and Systems (IROS)},
  pages={7493--7500},
  year={2020},
  organization={IEEE}
}

@article{ravi2024sam,
  title={Sam 2: Segment anything in images and videos},
  author={Ravi, Nikhila and Gabeur, Valentin and Hu, Yuan-Ting and Hu, Ronghang and Ryali, Chaitanya and Ma, Tengyu and Khedr, Haitham and R{\"a}dle, Roman and Rolland, Chloe and Gustafson, Laura and others},
  journal={arXiv preprint arXiv:2408.00714},
  year={2024}
}

@article{sapkota2025yolo26,
  title={YOLOv8: key architectural enhancements and performance benchmarking for real-time object detection},
  author={Sapkota, Ranjan and Cheppally, Rahul Harsha and Sharda, Ajay and Karkee, Manoj},
  journal={arXiv preprint arXiv:2509.25164},
  year={2025}
}

@inproceedings{akiba2019optuna,
  title={Optuna: A next-generation hyperparameter optimization framework},
  author={Akiba, Takuya and Sano, Shotaro and Yanase, Toshihiko and Ohta, Takeru and Koyama, Masanori},
  booktitle={Proceedings of the 25th ACM SIGKDD international conference on knowledge discovery \& data mining},
  pages={2623--2631},
  year={2019}
}

@inproceedings{raphalen2025haptic,
  title={Haptic shared control of a pair of microrobots for telemanipulation using constrained optimization},
  author={Raphalen, Leon and Ferro, Marco and Misra, Sarthak and Giordano, Paolo Robuffo and Pacchierotti, Claudio},
  booktitle={2025 IEEE/RSJ International Conference on Intelligent Robots and Systems (IROS)},
  pages={17391--17397},
  year={2025},
  organization={IEEE}
}

@article{zhang2025scheduling,
  title={Scheduling Adaptive Imitation Learning for Long-Horizon Dexterous Robot Micromanipulation of Deformable Cell},
  author={Zhang, Youchao and Shen, Xufang and Wang, Chuhan and Wang, Fanghao and Zhao, Antian and Lyu, Yining and Knoll, Alois and Liu, Ying and Ying, Yibin and Zhou, Mingchuan},
  journal={IEEE Robotics and Automation Letters},
  volume={11},
  number={1},
  pages={41--48},
  year={2025},
  publisher={IEEE}
}

\end{document}